\documentclass[12pt]{article}
\usepackage{graphicx}
\usepackage{amssymb}
\usepackage{amscd}
\usepackage{amsmath}
\usepackage{cite}
\usepackage{textcomp} 
\usepackage{float}

\begin{document}
\begin{titlepage}

{\hbox to\hsize{\hfill December 2018 }}

\bigskip \vspace{3\baselineskip}

\begin{center}
{\bf \Large 
More stringent constraints on the unitarised \\ fermionic dark matter Higgs portal}

\bigskip

\bigskip

{\bf Shyam Balaji and Archil Kobakhidze \\ }

\smallskip

{ \small \it
ARC Centre of Excellence for Particle Physics at the Terascale, \\
School of Physics, The University of Sydney, NSW 2006, Australia, \\
E-mails: shyam.balaji, archil.kobakhidze@sydney.edu.au 
\\}

\bigskip
 
\bigskip

\bigskip

{\large \bf Abstract}

\end{center}
\noindent 
We revisit the simplest model of Higgs portal fermionic dark matter. The dark matter in this scenario is thermally produced in the early universe due to the interactions with the Higgs boson which is described by a non-renormalisable dimension-5 operator. The dark matter-Higgs scattering amplitude grows as $\propto \sqrt{s}$, signalling a breakdown of the effective description of the Higgs-dark matter interactions at large enough (compared to the mass scale $\Lambda$ of the dimention-5 operator) energies. Therefore, in order to reliably compute Higgs-dark matter scattering cross sections, we employ the K-matrix unitarisation procedure. To account for the desired dark matter abundance, the unitarised theory requires appreaciably smaller $\Lambda$ than the non-unitarised version, especially for dark matter masses around and below the Higgs resonance, $m_{\chi}\lesssim 65$ GeV, and $m_{\chi}\gtrsim $ few TeV. Consequently, we find that the pure scalar CP-conserving model is fully excluded by current direct dark matter detection experiments.

 \end{titlepage}

\section{Fermionic Higgs portal model}

The model-independent treatment of dark matter within the effective field theory (EFT) formalism is severely constrained due to the applicability of EFT in high energy regimes. In \cite{Bell:2016obu}, collaborators have previously advocated utilisation of the K-matrix unitarisation procedure \cite{Chung:1995dx} to remedy the problem. Unitararisation represents a tool for theoretically reliable calculations of cross sections and other observables of interest in the domain where the EFT is usually assumed inapplicable. For example, the standard EFT description is not appropriate for collider dark matter searches at high energies, while unitarised theories give sensible results and in certain cases, are better suited for data analysis \cite{Bell:2016obu}  in comparison to the truncation method shown in \cite{Busoni:2013lha}. Furthermore, naive EFT calculations may lead to a substantially lower dark matter abundance due to the overestimation of the thermally-averaged dark matter annihilation cross section, which can be improved significantly via the unitarised EFT. In this work we demonstrate usefulness of the K-matrix unitarisation for the simplest fermionic Higgs portal dark matter model \cite{Kim:2014, Fedderke:2014wda}. In particular, we demonstrate that the unitarised CP-conserving Higgs-dark matter interactions are actually fully excluded by the latest direct detection data. This is in contrast with recent results in \cite{Athron:2018vxy}, where non-unitarised cross-sections were used.   

We hypothesise a dark matter Dirac fermion\footnote{The case of Majorana fermion can be treated analogously and the results are not significantly different.}, $\chi$,  of mass $m_{\chi}$. It carries no Standard Model gauged charges and, thus, the lowest order effective operator that describes dark matter interactions with the Standard Model particles is
\begin{align}
\mathcal{L} &=\frac{1}{\Lambda}H^{\dagger}H \bar\chi(\cos\xi+i\gamma_5 \sin\xi)\chi  \nonumber \\
& =\frac{1}{\Lambda}\left(v h+\frac{1}{2}h^2\right)\bar\chi(\cos\xi+i\gamma_5 \sin\xi)\chi~,
\label{1}
\end{align}
where $\Lambda$ is the EFT cut-off scale parameter and $\xi$ is the CP-violating phase. In the second line we explicitly expanded the electroweak Higgs doublet $H$ around its expectation value $v\approx246$ GeV in the unitary gauge, $H=2^{-1/2}(0,v+h)^{\rm T}$ and the Higgs condensate contribution to the dark matter mass is absorbed in $m_{\chi}$. 

In the low energy domain of the theory, $E<< \Lambda$, the Higgs-dark matter portal (\ref{1}) is dominated by the relevant dimension-4 $h\chi\bar\chi$ operators. These operators are renormalisable and also perturbative, provided $v\lesssim \Lambda$, and hence calculations in this domain of the theory are under control. However, the high energy dark matter-Higgs interactions are dominated by non-renormalisable $h^2\chi\bar\chi$ interactions. In fact, for $E\gtrsim\Lambda$ the scattering amplitudes described by $h^2\chi\bar\chi$ operators grow as $E/\Lambda$, signalling violation of perturbative unitarity. This violation of quantum-mechanical unitarity is actually fictitious and reflects inapplicability of perturbative treatment to the effective non-renormalisable interactions. 

In the context of thermal dark matter considered here, violation of perturbative unitarity has a two-fold effect on dark matter phenomenology. First, the naive application of the theory in the 'forbidden' energy domain, leads to overestimation of thermally averaged cross sections, which are then used to estimate dark matter abundance. Hence, more suppressed (larger values of $\Lambda$) Higgs-dark matter interactions are required to fit the observed abundance in this approach. Second, a significant area in parameter space, associated with high energy processes, is simply declared theoretically intractable. Therefore,  no reliable physical information can be extracted, even if experimental data are available to probe that region of parameters. As will be shown below, the unitarisation procedure is a remedy to these problems.

\section{K-matrix unitarisation of the fermion dark matter Higgs portal}
It is useful to recall K-matrix unitarisation formalism in a general context first. The unitarity of scattering operator $S$
\begin{equation}
S=1+2 i T~,
\label{K-1}
\end{equation} 
implies that the transition operator $T$ satisfies the following constraint (the optical theorem)
\begin{equation}
T-T^{\dagger}=2 iT^{\dagger}T~.
\label{K-2}
\end{equation} 
We define the $K$ operator as the solution of the equations
\begin{equation}
K=T-i TK~.
\label{K-3}
\end{equation}
If one regards $K$ as known with $T$ solved from Eq. (\ref{K-3}), then $T$ will satisfy the unitarity constraint Eq. (\ref{K-2}) if and only if $K$ is Hermitian, i.e. $K^{\dagger}=K$. Within perturbation theory, the expansion $T=T_0+T_1+...$, implies that one can approximate $K$ by the tree-level contribution $T_0$ to the full $T$-operator, i.e. $K=T_0$, providing $T_0$ is Hermitian\footnote{In fact, for time-reversal invariant scattering processes, $T_0$ is actually symmetric and real.}. If so, the unitarised tree-level $T$ operator can simply be written as
\begin{equation}
T^{U}=\frac{T_0}{1-i T_0}~,
\label{K-4}
\end{equation} 
 where $T_0^{\dagger}=T_0$ is assumed to be satisfied. For small $T_0$, $T^{U}\simeq T_0$, while for large $T_0$, the unitarised operator becomes $T^{U}=i$. Therefore, although $T^{U}$ does not reflect the potential existence of new resonances, it gives a reliable estimation of the process cross sections (at least away from the resonance energies), unlike the naive tree-level $T_0$ which will result in an overestimated cross section due to the violation of perturbative unitarity. For the purpose of this work, the simple unitarisation procedure described by Eq. (\ref{K-4}) turns out to be sufficient and is performed numerically as this is computationally efficient.
 
The generic  partial wave expansion for the $T$-matrix in the helicity basis can be written
\begin{align}
\label{eq:PWexpansion}
T^J_{\lambda'\lambda} = \textlangle J\lambda_c\lambda_d|T|J\lambda_a\lambda_b\textrangle = \int d\Omega \textlangle \Omega \lambda_c\lambda_d|T|0\lambda_a\lambda_b\textrangle D^J_{\lambda\lambda'}(\phi,\theta,0)~,
\end{align}
where $\lambda_a$ , $\lambda_b$ and $\lambda_c$ , $\lambda_d$ are the initial and final state particle helicities respectively. The other terms are defined $\lambda=\lambda_a-\lambda_b$ , $\lambda'=\lambda_c-\lambda_d$ and the Wigner D-functions are denoted $D^J_{\lambda\lambda'}(\phi,\theta,0)$. The partial wave expansion only consists of the terms with total angular momentum $J=0$. The others go to 0 when the Wigner D-functions (for $J=0$ this is simply $D^0_{\lambda\lambda'}=1$) are substituted and integrated over the solid angle. The $T$-matrix is related to the familiar Lorentz invariant amplitude $\mathcal{M}_{fi}$ by
\begin{align}
\label{eq:TmatrixandAmplitude}
\textlangle\Omega \lambda_c\lambda_d|T|0\lambda_a\lambda_b\textrangle = \frac{1}{32\pi^2}\sqrt{\frac{4 p_f p_i}{s}}\mathcal{M}_{fi}~,
\end{align}
where $p_f$ and $p_i$ are the  initial and final state particle momenta for $2\rightarrow2$ scattering in the CoM frame. The total non-averaged cross-section can then be written as:
\begin{align}
\sigma_{fi} = \frac{4\pi}{s-4m_i^2} \sum_{hel}\sum_{J} (2J+1) |T^J_{\lambda'\lambda}|^2~.
\end{align}
This cross-section has to be further averaged over initial state helicities. 

We apply this formalism to the fermionic dark matter Higgs portal model described by the Lagrangian (\ref{1}). The dark matter abundance is defined through the thermally averaged rate of dark matter particle scatterings in the early universe
\begin{align}
\langle\sigma v\rangle=\frac{1}{8 m_{\chi}^4 T K^2_{2}(m_\chi/T)}\int_{4m_\chi^2}^{\infty}\sigma(s)(s-4 m_{\chi}^2) \sqrt{s} K_1(\sqrt{s}/T)ds~,
 \label{A-1}
\end{align}
where $K_{1,2}(x)$ are modified Bessel functions of the second kind and $\sigma (s)$ is the total scattering cross section at energy $\sqrt{s}$. The inverse freeze out temperature $x_F=m_{\chi}/T$ is determined by the iterative equation.
\begin{align}
x_F=ln\left(\frac{m_\chi}{2\pi^3}\sqrt{\frac{45m_{pl}^2}{2 g_{*}x_F}}\langle\sigma v\rangle\right),
\end{align}
where $g_{*}$ counts the effective degrees of freedom in equilibrium and $m_{pl}$ is the Planck mass. Within the EFT non-renormalisable model (\ref{1}), this cross section computed using standard Feynman rules is not reliable for large $\sqrt{s}>>\Lambda$.  In particular, the tree-level $J=0$ partial waves (see the Appendix A for notations and conventions for helicity eigenstates and scattering matrix elements) for $\chi \bar{\chi} \leftrightarrow hh$ scatterings actually diverge at large $\sqrt{s}$
\begin{align}
T_{0~\chi_{L,R} \bar{\chi}_{L,R} \rightarrow hh} = \pm\frac{((s-4m_h^2)(s-4m_\chi^2))^{\frac{1}{4}}(\sqrt{s-4m_\chi^2}\cos\xi\mp i\sqrt{s}\sin\xi)}{8\pi \sqrt{s}\Lambda}
\longrightarrow \propto \frac{\sqrt{s}}{8\pi\Lambda}~,
\label{A-2}
\end{align}
and also $T_{0~hh \rightarrow \chi_{L,R} \bar{\chi}_{L,R}} =T^{*}_{0~\chi_{L,R} \bar{\chi}_{L,R} \rightarrow hh}$. Similarly,  tree-level scattering amplitudes $\chi \bar{\chi} \leftrightarrow VV$ involving the longitudinal  electroweak gauge bosons ($V\equiv W^{\pm}, Z^0$) behave as
\begin{align}
T_{0~\chi_{L,R} \bar{\chi}_{L,R} \rightarrow VV} =\mp \frac{(2m_V^2-s)((s-4m_V^2)(s-4m_\chi^2))^{\frac{1}{4}}\left(\sqrt{s-4m_\chi^2}\cos\xi\mp i\sqrt{s}\sin\xi\right)}{8\pi \sqrt{s}\Lambda (s-m_h^2+im_h\Gamma_h)}\longrightarrow \propto \frac{\sqrt{s}}{8\pi \Lambda}~,
\label{A-3}
\end{align}
and $T_{0~VV \rightarrow \chi_L \bar{\chi}_L} =T^{*}_{0~\chi_L \bar{\chi}_L \rightarrow VV}$, $T_{0~VV \rightarrow \chi_R \bar{\chi}_R} =T^{*}_{\chi_R \bar{\chi}_R \rightarrow VV}$ in the limit where $\Gamma_h << m_h$. Therefore, the total cross section $\sigma(s)$ computed within EFT is not reliable at large values of $s$. Although the integrand in Eq. (\ref{1}) is Boltzmann suppressed, we find that the resulting thermal averaged cross section is still high, so that relatively large values of $\Lambda$ are required to account for the observed dark matter abundance.   

We compute the unitarised thermally averaged dark matter cross section including the relevant $2\rightarrow2$ annihilation channels, $\chi\bar{\chi} \leftrightarrow \chi\bar{\chi}, VV, hh, f\bar{f}$ where  $f=t, b, c, \tau$  by employing the unitarisation prescription shown in Eq. (\ref{K-4}). Next, we fix the dark matter abundance and numerically compute the allowed parameter $\Lambda$ for a given dark matter mass $m_{\chi}$. This sets the model ready for direct confrontation against current experimental limits. The required strength of the portal interactions [$\sim 1/\Lambda$, see Eq. (\ref{1})] is larger in the unitarised theory compared to the non-unitary one and, especially for dark matter masses around and below the Higgs resonance, $m_{\chi}\lesssim 62.5$ GeV, and $m_{\chi}\gtrsim $ few TeV. Consequently, we obtain more stringent constraints as detailed below.

\section{Results}
\begin{figure}[h]
\begin{center}
\includegraphics[width=8cm]{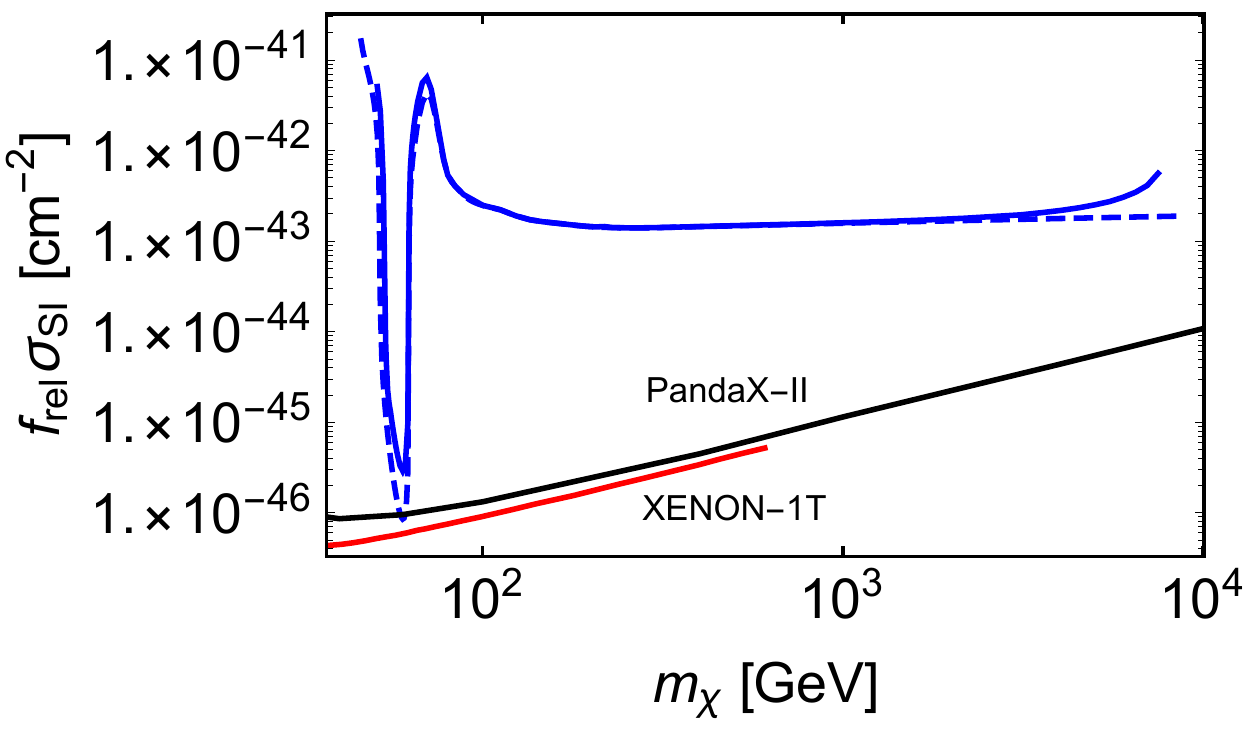}~~
\includegraphics[width=8cm]{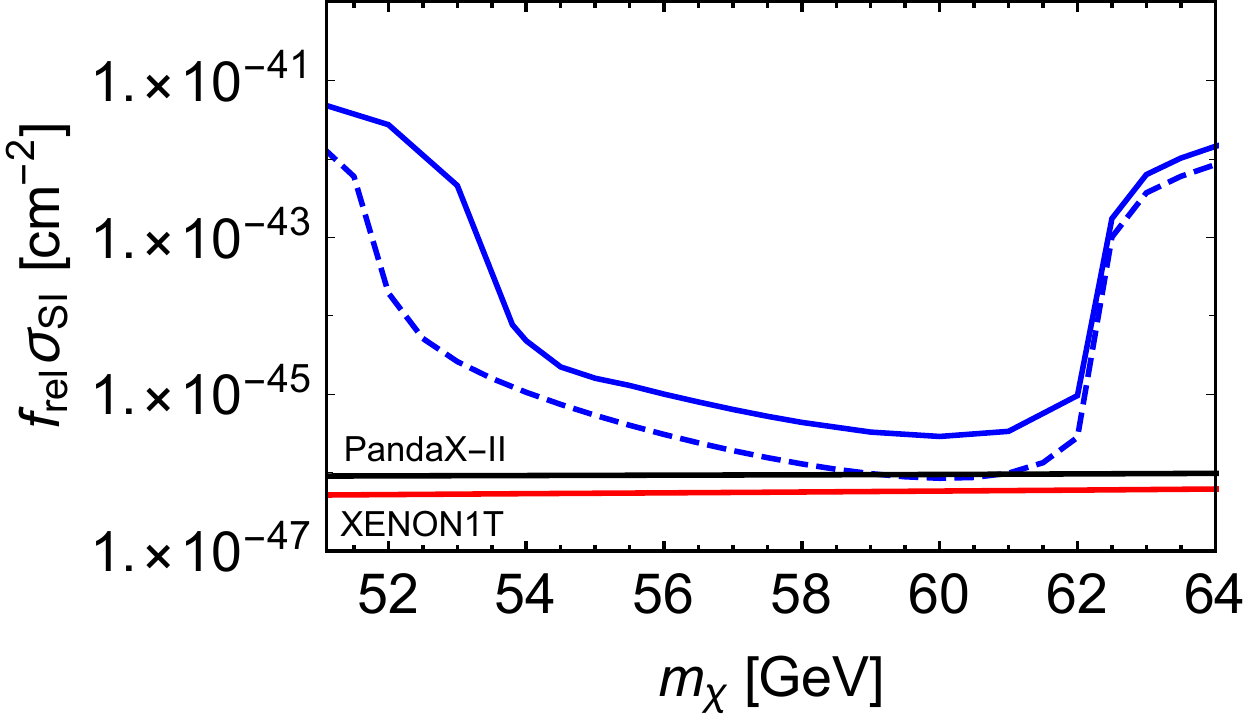} \\ \vspace{0.5cm}
\includegraphics[width=8cm]{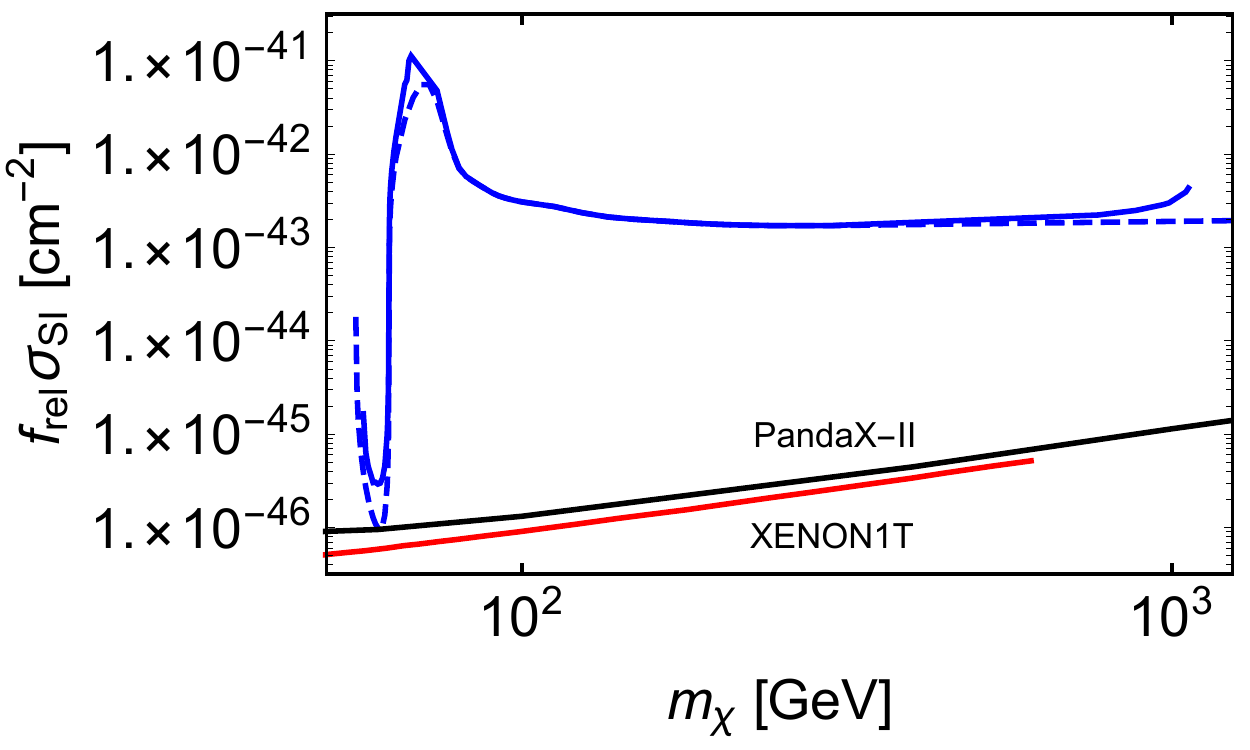} ~~ 
\includegraphics[width=8cm]{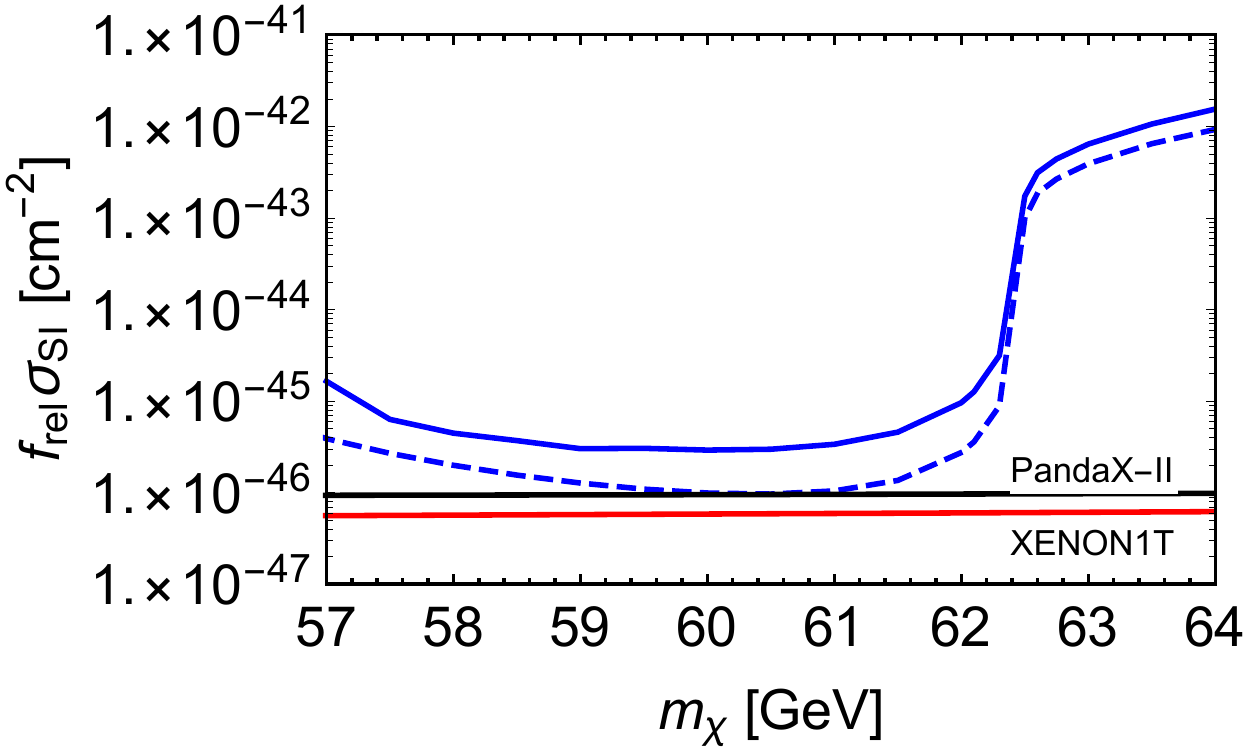}  
\end{center} 
\vspace{-0.7cm}
\caption{\small The rescaled spin-independent nucleon-dark matter cross section as a function of dark matter mass within the fermion dark matter Higgs portal model in the pure scalar channel, $\xi=0$. The two upper plots correspond to the dark matter fermion $\chi$ constituting 100\% ($f_{rel}=1$) of dark matter, while the graphs on the lower panels correspond to $\chi$ being 10\% ($f_{rel}=0.1$) fraction of the total dark matter.  A small range of masses (see the left panels) are still allowed by the latest PandaX-II \cite{Cui:2017nnn} and, within $2\sigma$, by XENON1T  \cite{Aprile:2018dbl} data in the non-unitarised theory (dashed curve). However,  the fermionic dark matter in the scalar channel is fully excluded in the unitarised theory (solid curve).}
\label{f1} 
\end{figure}

\begin{figure}[h]
\begin{center}
\includegraphics[width=8cm]{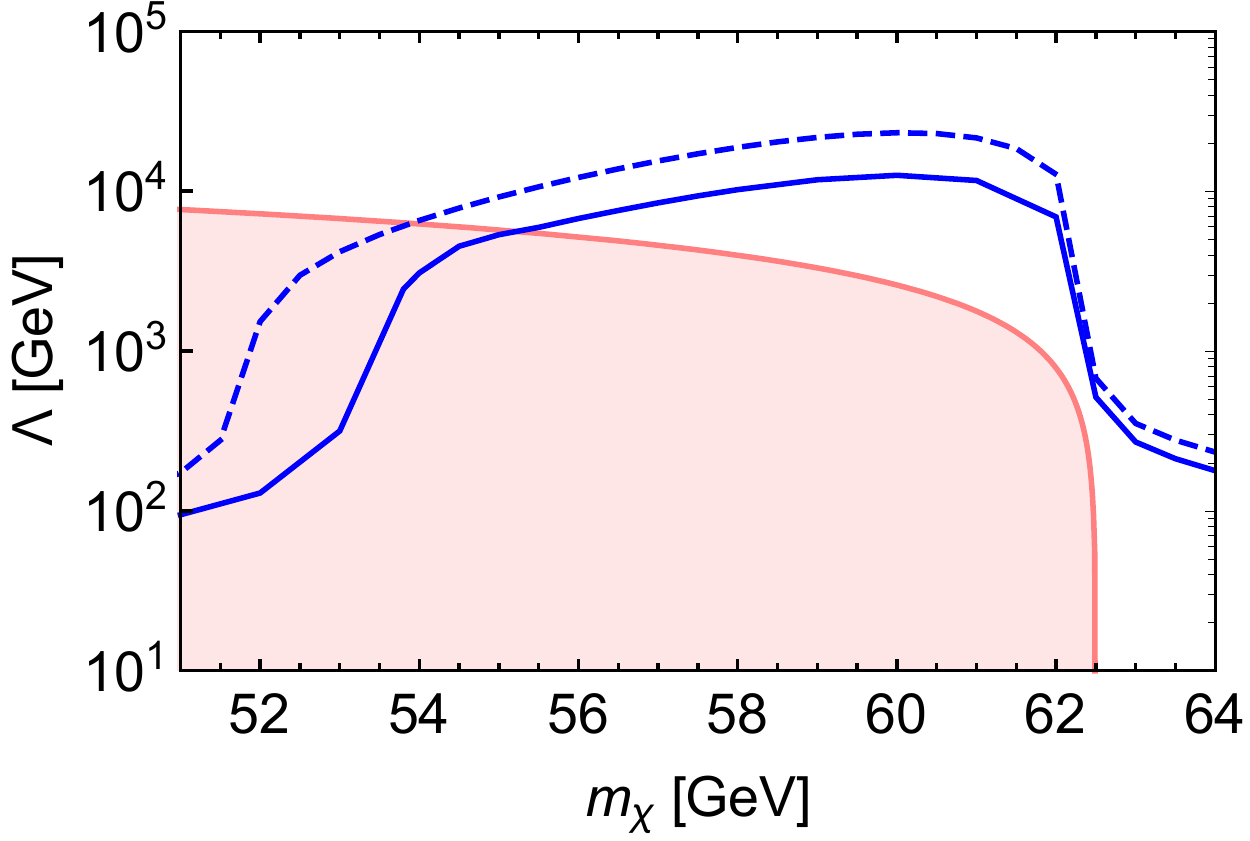}~~
\includegraphics[width=8cm]{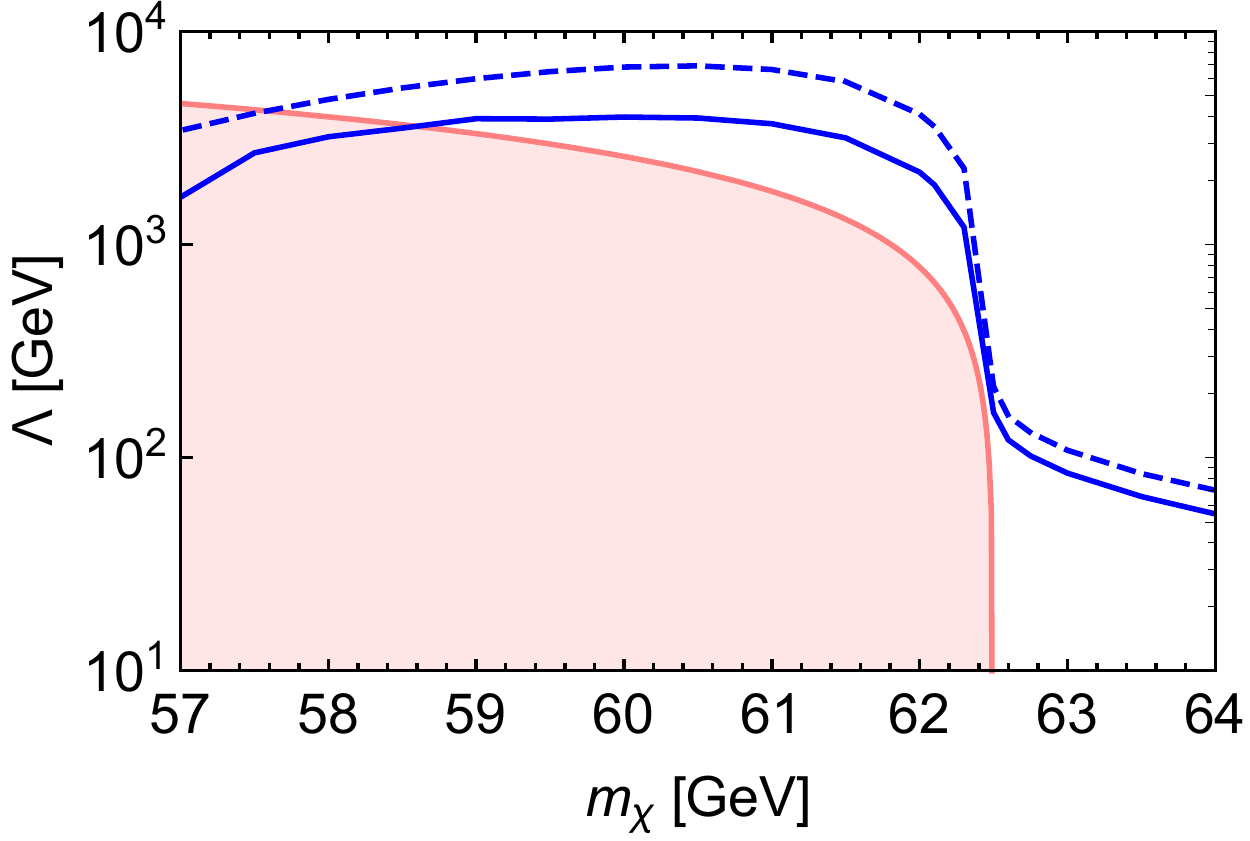} \\ \vspace{0.5cm}
\includegraphics[width=8cm]{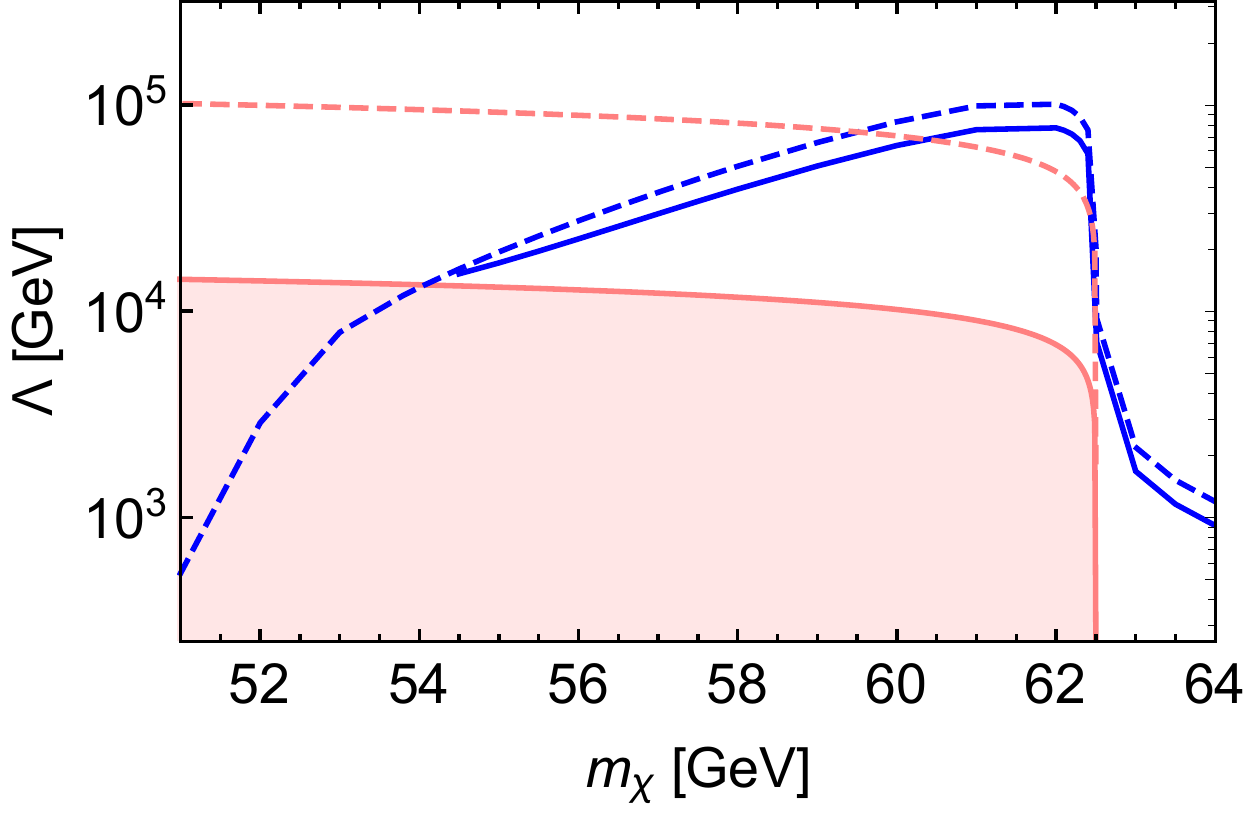} ~~ 
\includegraphics[width=8cm]{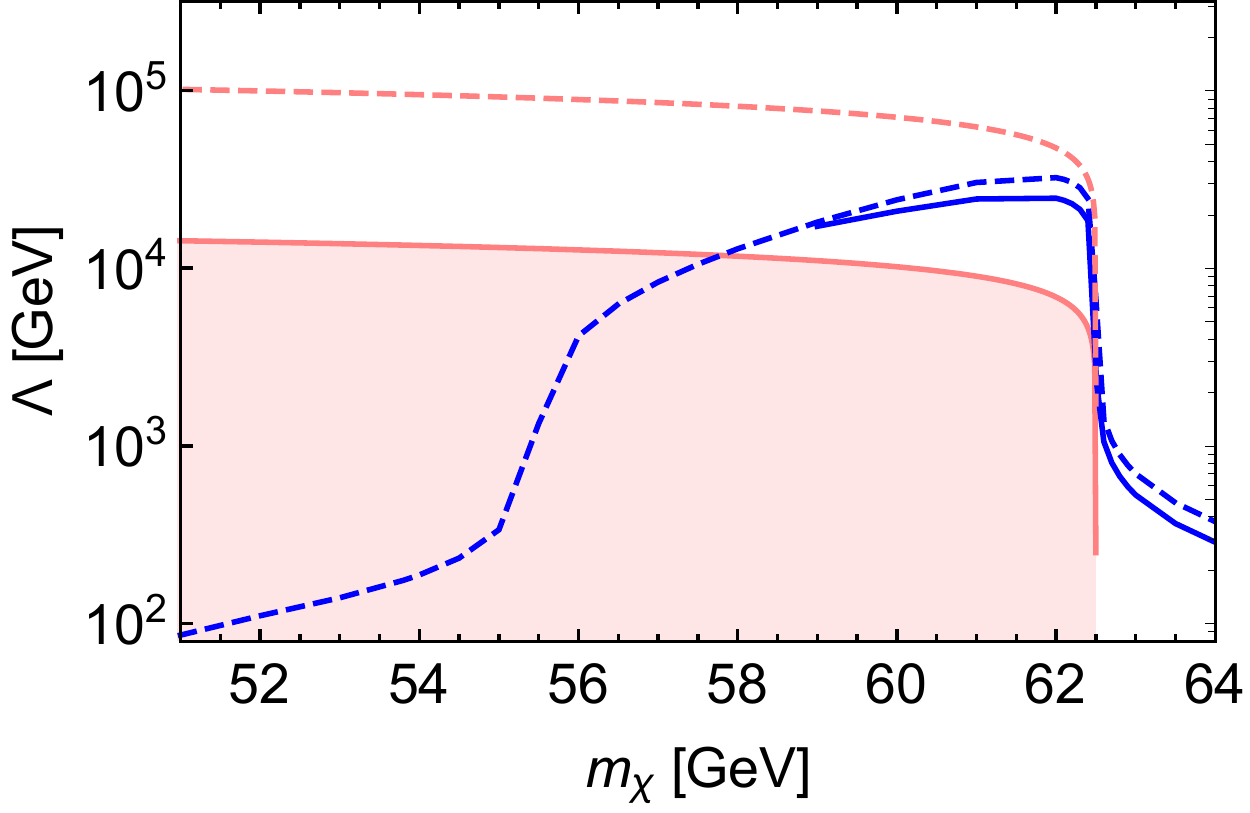}
\end{center} 
\vspace{-0.7cm}
\caption{\small Constraints (shaded regions) on the fermionic dark matter Higgs portal models in the pure scalar, $\xi=0$ (upper panels), and pure pseudoscalar, $\xi=\pi/2$ (lower panels) channels from the LHC measurements of the Higgs invisible width \cite{Aad:2015pla, Khachatryan:2016whc} (see the exclusion region below the solid curve  $BR(h\rightarrow\chi\bar{\chi})<0.19$). The left panels correspond to 100\% fraction of dark matter, while the right panels correspond to the 10\%  fraction of dark matter. The pseudoscalar channel can be potentially excluded by measurements of the Higgs invisible width at ILC with the projected accuracy \cite{Asner:2013psa} (see the exclusion region below the dashed curve $BR(h\rightarrow\chi\bar{\chi})<0.004$).}
\label{f2} 
\end{figure}

The results of our analyses are presented graphically on Figures \ref{f1} and \ref{f2}. The solid curves on these figures correspond to the dark matter abundance computed within the unitarisation formalism discussed in the previous section, while dashed curves correspond to dark matter abundance in the non-unitarised theory. We have explicitly considered two limiting cases of pure scalar, $\xi=0$, and pure pseudoscalar, $\xi=\pi/2$, channels of the Higgs-dark matter portal, Eq. (\ref{1}).

From Figure \ref{f1} we observe that in the non-unitarised theory the pure scalar channel within 2$\sigma$ is compatible with the latest XENON1T data \cite{Aprile:2018dbl} in a very narrow range of dark matter masses (the so-called "resonant Higgs portal"), $m_{\chi}\approx 59-61$ GeV (see also \cite{Athron:2018vxy}). In stark contrast, within the unitarised theory, the pure scalar channel is fully excluded. We should also note, that the non-unitarised theory is not applicable for large dark matter masses, $m_{\chi}>4\pi\Lambda$, and is hence formally unconstrained. This range of masses is now also excluded in the unitarised theory. 

The light dark matter with masses $m_{\chi}<m_h/2$ is also constrained from the LHC data \cite{Aad:2015pla, Khachatryan:2016whc} on invisible Higgs decays (see Figure \ref{f2}). These constraints are particularly important for the pseudoscalar Higgs portal, to which direct search experiments are not sensitive due to the velocity suppression of the nucleon-dark matter cross section (\ref{wdm}). The current data allow  masses $m_{\chi}\approx 55-62.5$ GeV and  $m_{\chi}\approx 58-62.5$ GeV for 100\% and 10\% dark matter fractions, respectively. The pseudoscalar channel can be further constrained/excluded by measurements of the Higgs invisible branching ratio with $0.4$\% projected accuracy at the ILC.

\section{Conclusion}

We have revisited the fermionic dark matter Higgs portal EFT by applying the K-matrix unitarisation formalism. Within the unitarised EFT the relevant scattering processes can be computed reliably in the entire energy range. Consequently, we were able to obtain more reliable theoretical thermally average dark matter cross sections and hence the dark matter abundance. By fixing the desired dark matter abundance we computed the cut-off scale parameter $\Lambda$, which turns out to be lower compared to the one obtained in the non-unitarised theory. Furthermore, unlike the non-unitarised theory, the unitarised EFT is also applicable for heavy dark matter masses, $m_{\chi}\geq4\pi \Lambda$, hence the constraints can be extended to that parameter range as well. 

We have found that the fermionic dark matter in the pure scalar channel is fully excluded by recent direct dark matter search experiments \cite{Aprile:2018dbl, Cui:2017nnn}. For the pure pseudoscalar channel the most stringent constraints come from the Higgs invisible decay width measurements at LHC \cite{Aad:2015pla, Khachatryan:2016whc} for $m_{\chi}\leq m_h/2$. We again found more stringent (albeit marginally) constraints in the unitarised theory. This range of parameters can be largely excluded by precision measurements of the Higgs invisible decay width at ILC. 

The unitarised EFT formalism is a powerful theoretical tool to analyse dark matter search experiments in a model-independent way, the range of parameters and energies where the usual EFT approach fails. This also concerns a number of simplified models without manifest gauge invariance. In the context of Higgs portal models, the massive vector dark matter model is expected to be more constrained within the unitarised theory.

\paragraph{Acknowledgements} We would like to thank Michael Schmidt and Pat Scott for useful discussions. The work was partly supported by the Australian Research Council. AK is also supported by the Rustaveli National Science Foundation. 

\appendix
\section{Helicity formalism and matrix elements}
We must first define the helicity eigenstates before computing the relevant process matrix elements. Helicity spinors for fermions are defined as per \cite{Thomson:2013} (where $\theta$ and $\phi$ are defined as polar and azimuthal angles respectively) and are listed below

\begin{align}
 u_R =\sqrt{E+m} \begin{pmatrix}
 \cos\frac{\theta}{2}\\ 
e^{i\phi}\sin\frac{\theta}{2}\\
\frac{\sqrt{E^2-m^2} \cos\frac{\theta}{2}}{E+m}\\
 \frac{\sqrt{E^2-m^2} e^{i\phi} \sin\frac{\theta}{2}}{E+m}\end{pmatrix} &,
u_L =\sqrt{E+m}  \begin{pmatrix}
 -\sin\frac{\theta}{2}\\ 
e^{i\phi}\cos\frac{\theta}{2}\\
\frac{\sqrt{E^2-m^2} \sin\frac{\theta}{2}}{E+m}\\
 -\frac{\sqrt{E^2-m^2} e^{i\phi} \cos\frac{\theta}{2}}{E+m}\end{pmatrix} &, \\
 v_R =\sqrt{E+m} \begin{pmatrix}
\frac{\sqrt{E^2-m^2} \sin\frac{\theta}{2}}{E+m}\\
 -\frac{\sqrt{E^2-m^2} e^{i\phi} \cos\frac{\theta}{2}}{E+m}\\
 -\sin\frac{\theta}{2}\\ 
e^{i\phi}\cos\frac{\theta}{2}\end{pmatrix} &,
v_L =\sqrt{E+m} \begin{pmatrix}
\frac{\sqrt{E^2-m^2} \cos\frac{\theta}{2}}{E+m}\\
 \frac{\sqrt{E^2-m^2} e^{i\phi} \sin\frac{\theta}{2}}{E+m}\\
 \cos\frac{\theta}{2}\\ 
e^{i\phi}\sin\frac{\theta}{2}
\end{pmatrix}~. 
\end{align}
It should be noted that $L$ and $R$ here denote helicity and not chirality polarizations. Polarization vectors satisyfing transversality relations for the gauge bosons can be written as usual
\begin{align}
k^\mu =\begin{pmatrix}E\\k_x\\k_y\\k_z\end{pmatrix},  
\varepsilon_{1}^{\mu}(k) =\frac{1}{|\vec{k}|k_T}\begin{pmatrix}0\\k_x k_z\\k_y k_z\\-k_T^2\end{pmatrix},   
\varepsilon_{2}^{\mu}(k) =\frac{1}{k_T}\begin{pmatrix}0\\-k_y\\k_x\\0\end{pmatrix}, 
\varepsilon_{3}^{\mu}(k) =\frac{E}{m|\vec{k}|}\begin{pmatrix}\frac{|\vec{k}|^2}{E}\\k_x\\k_y\\k_z\end{pmatrix}~, 
\end{align}
where
\begin{align}
k_T =\sqrt{k_x^2+k_y^2}~.
\end{align}
The normalized transverse helicity eigenvectors can now be written as:
\begin{align}
\varepsilon_{\pm}^{\mu}(k) =\frac{1}{\sqrt{2}}(\mp \varepsilon_1^\mu(k)-i \varepsilon_2^\mu(k))~,
\end{align}
and for massive vector bosons we have the longitudinal polarization vector given by:
\begin{align}
\varepsilon_{L}^{\mu}(k) =\varepsilon_3^\mu(k)~.
\end{align}

We may now turn our attention to the the computation of the explicit leading order $J=0$ partial waves for each process. The matrix elements for $\chi \bar{\chi} \rightarrow \chi \bar{\chi}$ with both $s$ and $t$ channel scattering via a Higgs mediator which yields
\begin{align}
\mathcal{M}_{\chi_L \bar{\chi}_L \rightarrow \chi_L \bar{\chi}_L} = \frac{v^2}{\Lambda^2} \left(-\frac{4 m_\chi^2 \cos^2\xi (1+\cos\theta)}{2m_h^2+(s-4m_\chi^2)(\cos\theta-1)-2im_h\Gamma_h}+\frac{2 m_\chi^2-s+2m_\chi^2 \cos2\xi}{s-m_h^2+i m_h \Gamma_h}\right)~,
\end{align}
\begin{align}
\mathcal{M}_{\chi_L \bar{\chi}_L \rightarrow \chi_R \bar{\chi}_R} = \frac{1}{\Lambda^2} \left(\frac{\left(\sqrt{s-4m_\chi^2}v \cos\xi -i \sqrt{s} v \sin\xi \right)^2}{s-m_h^2+i m_h \Gamma_h}+\frac{2 v^2 \sin^2\frac{\theta}{2}\left(i\sqrt{s}\cos\xi+\sqrt{s-4 m_\chi^2}\sin\xi\right)^2}{2m_h^2+(s-4m_\chi^2)(1-\cos\theta)-2im_h\Gamma_h}\right)~,
\end{align}
\begin{align}
\mathcal{M}_{\chi_R \bar{\chi}_R \rightarrow \chi_L \bar{\chi}_L} = \frac{1}{\Lambda^2} \left(\frac{\left(\sqrt{s-4m_\chi^2}v \cos\xi +i \sqrt{s} v \sin\xi \right)^2}{s-m_h^2+i m_h \Gamma_h}-\frac{2 v^2 \sin^2\frac{\theta}{2}\left(\sqrt{s}\cos\xi+i\sqrt{s-4 m_\chi^2}\sin\xi\right)^2}{2m_h^2+(s-4m_\chi^2)(1-\cos\theta)-2im_h\Gamma_h}\right)~,
\end{align}
with $\mathcal{M}_{\chi_R \bar{\chi}_R \rightarrow \chi_R \bar{\chi}_R} = \mathcal{M}_{\chi_L \bar{\chi}_L \rightarrow \chi_L \bar{\chi}_L}$.

The matrix elements for $\chi \bar{\chi} \rightarrow hh$ processes read:
\begin{align}
\mathcal{M}_{\chi_L \bar{\chi}_L \rightarrow hh} = \frac{\sqrt{s-4m_\chi^2}\cos\xi-i\sqrt{s}\sin\xi}{\Lambda},
\end{align}
\begin{align}
\mathcal{M}_{\chi_R \bar{\chi}_R \rightarrow hh} = -\frac{\sqrt{s-4m_\chi^2}\cos\xi+i\sqrt{s}\sin\xi}{\Lambda},
\end{align}
with $\mathcal{M}_{hh \rightarrow \chi_L \bar{\chi}_L} =\mathcal{M}^{*}_{\chi_L \bar{\chi}_L \rightarrow hh}$ and $\mathcal{M}_{hh \rightarrow \chi_R \bar{\chi}_R} =\mathcal{M}^{*}_{\chi_R \bar{\chi}_R \rightarrow hh}$. 

The leading order tree-level scattering processes involving the longitudinal degrees of freedom (relevant in the large $s$ limit where pertubative unitarity is violated) of the electro-weak gauge bosons (where $VV$ can either be $W^{+}W^{-}$ or $Z^0Z^0$) are given by
\begin{align}
\mathcal{M}_{\chi_L \bar{\chi}_L \rightarrow VV} =- \frac{\left(s-2m_V^2\right)\left(\sqrt{s-4m_\chi^2}\cos\xi-i\sqrt{s}\sin\xi\right)}{\Lambda(s-m_h^2+im_h\Gamma_h)}~,
\end{align}
\begin{align}
\mathcal{M}_{\chi_R \bar{\chi}_R \rightarrow VV} = \frac{\left(s-2m_V^2\right)\left(\sqrt{s-4m_\chi^2}\cos\xi+i\sqrt{s}\sin\xi\right)}{\Lambda(s-m_h^2+im_h\Gamma_h)}~,
\end{align}
and $\mathcal{M}_{VV \rightarrow \chi_L \bar{\chi}_L} ={M}^{*}_{\chi_L \bar{\chi}_L \rightarrow VV}$, $\mathcal{M}_{VV \rightarrow \chi_R \bar{\chi}_R} ={M}^{*}_{\chi_R \bar{\chi}_R \rightarrow VV}$ in the limit where $\Gamma_h << m_h$. Note that amplitudes for $\chi\chi \leftrightarrow hh, VV$ processes diverge as $\sqrt{s}$ for large $s$.

Finally, the leading order tree-level scattering processes to generic final state SM fermions $f$ occur via the $s$-channel exchange of a Higgs boson and are given by
\begin{align}
\mathcal{M}_{\chi_L \bar{\chi}_L \rightarrow f_L \bar{f}_L} = \frac{m_f \sqrt{s-4m_f^2}\left(-\sqrt{s-4m_f^2}\cos\xi+i\sqrt{s}\sin\xi\right)}{\Lambda(s-m_h^2+im_h\Gamma_h)}~,
\end{align}
\begin{align}
\mathcal{M}_{\chi_R \bar{\chi}_R \rightarrow f_L \bar{f}_L} = \frac{m_f \sqrt{s-4m_f^2}\left(\sqrt{s-4m_f^2}\cos\xi+i\sqrt{s}\sin\xi\right)}{\Lambda(s-m_h^2+im_h\Gamma_h)}~,
\end{align}
with $\mathcal{M}_{\chi_L \bar{\chi}_L \rightarrow f_R \bar{f}_R} = -\mathcal{M}_{\chi_L \bar{\chi}_L \rightarrow f_L \bar{f}_L} $ and $\mathcal{M}_{\chi_R \bar{\chi}_R \rightarrow f_R \bar{f}_R} =-\mathcal{M}_{\chi_R \bar{\chi}_R \rightarrow f_L \bar{f}_L}$. Also, all the corresponding time-reversed processes are given by $\mathcal{M}_{  f_i\bar{f}_i \rightarrow \chi_i \bar{\chi}_i}=\mathcal{M}^{*}_{  \chi_i \bar{\chi}_i \rightarrow  f_i\bar{f}_i }$ and $\mathcal{M}_{  f_i\bar{f}_i \rightarrow \chi_j \bar{\chi}_j}=\mathcal{M}_{  \chi_i \bar{\chi}_i \rightarrow  f_j\bar{f}_j }$ where $i\neq j$ in the limit where $\Gamma_h << m_h$ which is the case in the parameter regions explored in this work. In our case, we consider only the heaviest SM fermions where $f=t, b, c, \tau$ which are the dominant thermal processes of interest for the dark matter relic density calculation. From all of these processes, the $T$-matrix can be computed and written in the basis $|\chi_L\chi_L\textrangle$, $|\chi_R\chi_R\textrangle$, $|hh\textrangle$, $|W^{+} W^{-}\textrangle$, $|Z^{0} Z^{0}\textrangle$, $|t_L t_L\textrangle$, $|t_R t_R\textrangle$, $|b_L b_L\textrangle$, $|b_R b_R\textrangle$, $|c_L c_L\textrangle$, $|c_R c_R\textrangle$, $|\tau_L \tau_L\textrangle$ and $|\tau_R \tau_R\textrangle$. The full $T$-matrix expression can then be constructed from these amplitudes in this basis and then the $J=0$ partial wave can be computed with Eq. (\ref{eq:TmatrixandAmplitude}) and (\ref{eq:PWexpansion}) followed by the cross-section. The full $T$-matrix expression is ommitted here for the sake of brevity.

\section{Dark matter-nucleon cross section and Higgs invisible decay width}
For the sake of readers convenience, we provide below the dark matter-nucleon scattering cross section and the Higgs invisible decay width used in our calculations. 
 
The $t$-channel Higgs mediated elastic scattering of fermionic WIMP on nucleons spin-independent cross-section is given by
\begin{align}
\sigma^{\chi N}_{SI}=4.7 \times 10^{-38}cm^2 \left(\frac{m_\chi}{\Lambda}\right)^2\left(\frac{1GeV}{0.94GeV+m_\chi}\right)^2\left[\cos^2\xi+\frac{1}{2}\left(\frac{\mu_{\chi N}}{m_\chi}\right)^2\nu_\chi^2\right]
\label{wdm}
\end{align}
Where $\nu_\chi\sim220km/s$ is the DM speed in the nucleon's rest frame and $\mu_{\chi N}=\frac{m_\chi m_N}{m_\chi+m_N}$ is the reduced mass of the WIMP-nucleon system and $m_N$ is the nucleon mass.

The tree-level Higgs to invisible decay width is given by
\begin{align}
\Gamma_{h\rightarrow \bar{\chi}\chi}=\frac{m_h}{8\pi}\frac{\phi_{0}^2}{\Lambda^2}\sqrt{1-\frac{4 m_{\chi}^2}{m_h^2}}\left(1-\frac{4 m_{\chi}^2}{m_h^2}\cos^2\xi\right)
\end{align}
Where the total Higgs width is given by $\Gamma_h=\Gamma_{SM}+\Gamma_{h\rightarrow \bar{\chi}\chi}$ where $\Gamma_{SM}=4.21$MeV. This width will feature in the tree-level Higgs propagator when the Higgs is a mediator at tree-level.


\begin{thebibliography}{999}

\bibitem{Bell:2016obu}
  N.~Bell, G.~Busoni, A.~Kobakhidze, D.~M.~Long and M.~A.~Schmidt,
  JHEP {\bf 1608} (2016) 125
  doi:10.1007/JHEP08(2016)125
  [arXiv:1606.02722 [hep-ph]].
  
\bibitem{Chung:1995dx} For a review, see,  
  S.~U.~Chung, J.~Brose, R.~Hackmann, E.~Klempt, S.~Spanier and C.~Strassburger,
  Annalen Phys.\  {\bf 4} (1995) 404.
  doi:10.1002/andp.19955070504
  
\bibitem{Busoni:2013lha}
  G.~Busoni, A.~De Simone, E.~Morgante and A.~Riotto,
  Phys.\ Lett.\ B {\bf 728} (2014) 412
  doi:10.1016/j.physletb.2013.11.069
  [arXiv:1307.2253 [hep-ph]].
  
\bibitem{Kim:2014}
  Y.~G.~Kim, K.~Y.~Lee,
  JHEP {\bf 0805} (2008) 100
  doi:10.1088/JHEP08(2008)100
  [arXiv:0803.2932 [hep-ph]].

\bibitem{Fedderke:2014wda}
  M.~A.~Fedderke, J.~Y.~Chen, E.~W.~Kolb and L.~T.~Wang,
  JHEP {\bf 1408} (2014) 122
  doi:10.1007/JHEP08(2014)122
  [arXiv:1404.2283 [hep-ph]].
    
\bibitem{Athron:2018vxy}
  P.~Athron {\it et al.} [GAMBIT Collaboration],
  arXiv:1809.02097 [hep-ph].
  
\bibitem{Aprile:2018dbl}
  E.~Aprile {\it et al.} [XENON Collaboration],
  Phys.\ Rev.\ Lett.\  {\bf 121} (2018) no.11,  111302
  doi:10.1103/PhysRevLett.121.111302
  [arXiv:1805.12562 [astro-ph.CO]].
  
\bibitem{Cui:2017nnn}
  X.~Cui {\it et al.} [PandaX-II Collaboration],
  Phys.\ Rev.\ Lett.\  {\bf 119} (2017) no.18,  181302
  doi:10.1103/PhysRevLett.119.181302
  [arXiv:1708.06917 [astro-ph.CO]].

  
\bibitem{Aad:2015pla}
  G.~Aad {\it et al.} [ATLAS Collaboration],
  JHEP {\bf 1511} (2015) 206
  doi:10.1007/JHEP11(2015)206
  [arXiv:1509.00672 [hep-ex]].
  
\bibitem{Khachatryan:2016whc}
  V.~Khachatryan {\it et al.} [CMS Collaboration],
  JHEP {\bf 1702} (2017) 135
  doi:10.1007/JHEP02(2017)135
  [arXiv:1610.09218 [hep-ex]].

\bibitem{Asner:2013psa}
  D.~M.~Asner {\it et al.},
  arXiv:1310.0763 [hep-ph].

\bibitem{Thomson:2013} Thomson, M. 2013. {\em Modern Particle Physics.} UK: Cambridge University Press.

\end{thebibliography}
\end{document}